\documentstyle[epsfig]{aipproc}

\begin{document}

\title{Entropy, Macroscopic Information,
and Phase Transitions}

\author{Juan M.R. Parrondo\thanks{This work
has been financially supported by Direcci\'on General de
Ense\~nanza Superior e Investigaci\'on Cient\'{\i}fica T\'ecnica
(Spain) Project No. PB97-0076.}}
\address{Dep. F\'{\i}sica At\'omica,
Molecular y Nuclear. \\
Universidad Complutense de Madrid.
28040-Madrid, Spain}

\maketitle

\begin{abstract}
The relationship between entropy and information
is reviewed, taking into account that
information
is stored in macroscopic degrees of freedom, such
as the order parameter in a system exhibiting
spontaneous symmetry breaking. It is shown that most
problems of the relationship between entropy
and information, embodied in a variety of Maxwell demons,
are also present in any  symmetry breaking
transition.
\end{abstract}

\section{Introduction}
\label{sec:intro}

The relationship between entropy and information has
been subject of a long controversy, almost as old as
the Second Law of Thermodynamics.
The history of this controversy, closely linked
to the analysis of the Maxwell demon,
has been presented in detail in the excellent
book of reprints collected by Leff and Rex \cite{leff}.

Maxwell devised his demon to show the probabilistic
nature of the Second Law of Thermodynamics: a being
capable of measuring the position and velocity of the
molecules of a gas could in principle violate
the Second Law. Operating a
door in an adiabatic wall between two gases
at different temperatures, the demon could
induce a flow of energy
from the cold to the hot gas.
The conclusion is that information about
the microscopic details
of a system allows one to beat the Second Law.

\begin{figure}[b!]
\centerline{\epsfig{file=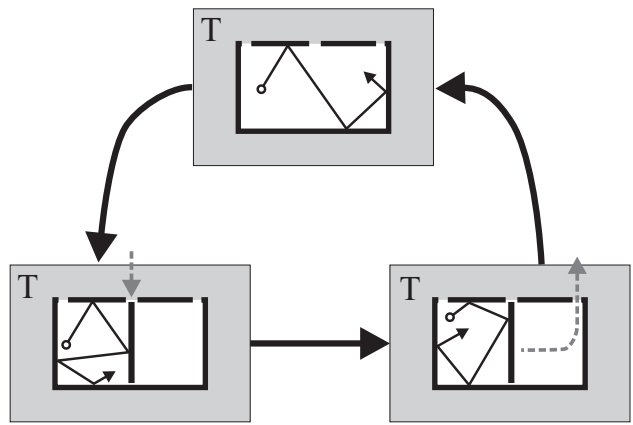,height=5.5cm}}
\vspace{10pt}
\caption{The Szilard engine.}
\label{szilard0}
\end{figure}

One  of the most relevant sequels of the
Maxwell demon is the Szilard engine \cite{leff,szil}.
It consists
of a box with a single-particle gas, i.e., a
particle that thermalizes
in any collision with the walls. A piston can be introduced (or
removed) either
at the middle of the box or at two opposite walls
(see Figure \ref{szilard0}).

The engine operates as follows. We insert the piston in the middle
of the box and {\em measure} in which side the particle gets
trapped. Then we let the gas expand quasistatically and remove the
piston. In the expansion the gas performs work:
\begin{equation}
\label{workszilard}
W= \int_V^{V/2} PdV =
kT\ln 2.
\end{equation}
This work can be used, for instance, to lift a weight
and store $kT\ln 2$ as potential energy.
The energy is taken from the thermal bath,
since the internal
energy of the gas is constant.
Therefore, the Szilard engine extracts energy
from a single thermal bath and performs work, in
contradiction with the Second Law of Thermodynamics.

Notice that, for the engine to work properly, it is absolutely
necessary to know in which side the particle gets trapped. In this
way, we can exert a pressure on the piston equal and opposite to
the pressure of the gas and let it expand quasistatically. On the
other hand, if the direction of the pressure were not correct, the
gas would expand irreversibly and Eq. (\ref{workszilard}) would
not hold. As in the original Maxwell demon, the Szilard engine can
beat the Second Law of Thermodynamics only if some information
about the state of the system is available.

The literature on the Szilard engine, as
well as on the Maxwell demon, has focused
mainly on the heat dissipation accompanying
the measurement, i.e., the acquisition of information,
and/or accompanying the erasure of this information
\cite{leff,szil,landauer,bennet,fahn}.
As an exception, Magnasco presented
in Ref. \cite{magszil} an interesting analysis
of the topology of the phase space of the engine.

Nevertheless, none of these papers has analyzed one of the obscure
points of the Szilard engine, namely, that it consists of
microscopic and macroscopic degrees of freedom interacting with
each other. This mixture of micro (the particle) and macro (the
piston) makes the Szilard engine a rather difficult and unclear
problem for many physicists, even for those working on Statistical
Mechanics.

In this paper I address this question, by giving
a novel interpretation to one of the steps
of the Szilard engine. The insertion of
the piston in the middle of the box can be
interpreted as a second order phase transition or
spontaneous symmetry breaking. The Hamiltonian
of the particle is symmetric under the
permutation of the two sides of the box.
However, the particle gets trapped
in only one of the sides.
This is equivalent to what happens in an Ising model
when it is driven from a paramagnetic to a ferromagnetic
phase in the absence of external magnetic field.

We will see below that all the astounding facts of the Szilard
engine are reproduced in the Ising model and in
any system exhibiting second order phase transitions.

The benefit of this new interpretation is twofold. On one side, it
helps to understand better the Szilard engine and the relationship
between entropy and information, since we will reach the same
conclusions without the use of single-particle gases interacting
with pistons. On the other side, it reveals that the consequences
of this relationship and the intriguing aspects of the Szilard
engine are not restricted to academic and artificial
constructions, such as the Maxwell demon and the Szilard engine
itself, but they are present in any spontaneous symmetry breaking,
that is to say, almost everywhere in Nature.

The paper is organized as follows. In Section \ref{sec:2procs}, I
analyze the energetics of two processes in the Szilard engine.
Section \ref{sec:sb} is a brief review of the concept of
spontaneous symmetry breaking and the Ising model. In Section
\ref{sec:processes}, I introduce two processes in the Ising model
which are equivalent to the processes studied in Section
\ref{sec:2procs}. Section \ref{sec:entro} discusses the
implications of the above results on the definition of entropy and
on the general validity of the Second Law. Finally, in Section
\ref{sec:open}, I present some conclusions and a list of open
problems.

\section{Two processes in the Szilard engine}
\label{sec:2procs}

\begin{figure}[b!]
\centerline{\epsfig{file=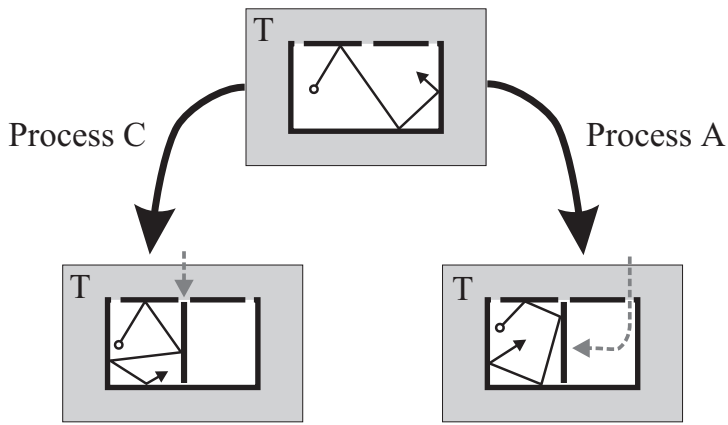,height=5.5cm}}
\vspace{10pt}
\caption{Processes $A$ and $C$ in the Szilard engine.}
\label{szilard1}
\end{figure}

Consider the Szilard gas and the processes $A$ and $C$ described
in Fig. \ref{szilard1}. In $C$, the piston is inserted in the
middle of the box and the particle gets trapped in one of the
sides. In $A$, the piston is introduced in the rightmost wall and
moved slowly to the middle of the box. Then, $C$ is the first step
of the Szilard cycle and $A$ is the inversion of the rest of the
cycle (cf. Figs. \ref{szilard0} and \ref{szilard1}).

Let us investigate the energetics of these two processes,
i.e., the energy transfer between the particle and
its surroundings. The particle exchanges energy with
two external systems: the thermal bath, and some
{\em external
agent} that handles the piston, exerting
pressure when it is needed. As in Thermodynamics,
I call {\bf heat}, $Q$, the energy transferred from
the thermal bath to the particle in a given
process and, {\bf work}, $W$,
the energy transferred from the system to
the external agent.
Finally, if $\cal U$ is the internal energy of the
particle, the First Law of Thermodynamics,
$ \Delta {\cal U} =Q-W $,
holds for any process.

In our particular case, process $C$
does not require
any work, or at least the work can be
arbitrarily small. On the other hand, process
$A$ involves a compression of the single-particle gas
to half of its volume and in this compression, if
 carried out quasistatically, a work $kT \ln 2$ is
done by the external agent.
Therefore, as defined above, work is given by:
\begin{equation}
W_A = -kT\ln 2\ ; \qquad
W_C = 0.
\label{wac}
\end{equation}
The internal energy of the particle remains
constant since the two processes are isothermal.
Thus, the heat in each process is:
\begin{equation}
Q_A = -kT\ln 2\ ; \qquad
Q_C = 0
\label{qac}
\end{equation}
i.e., along $A$, energy is transferred
from the system to the thermal bath.

The difference in the energetics of $A$ and $C$
is the key point of the Szilard engine.
The engine is nothing but the cycle $CA^{-1}$,
where $A^{-1}$ is the inverse of process $A$.
The energetics of $A^{-1}$ is $W_{A^{-1}}=-W_A$
and $Q_{A^{-1}}=-Q_A$,
only if $A^{-1}$ is the true inversion of $A$,
i.e., if the external agent
exerts a pressure equal to the pressure of the gas
and thus the expansion is done adiabatically.
In this case, we have $W_{CA^{-1}}=kT\ln 2$.

Notice that so far I have restricted the discussion to energy. The
consequences of the above results on the definition of entropy
will be explored in Section \ref{sec:entro}.

\section{Symmetry breaking transitions}
\label{sec:sb}

I have split the Szilard cycle into
processes $A$ and $C$, and showed that the
paradoxical nature of the engine lies
in the energetics of these two processes.

Moreover, as mentioned in the Introduction,
process $C$ can be seen as
a spontaneous symmetry breaking and
process  $A$ as a forced or non-spontaneous
symmetry breaking.
In fact,
symmetry breaking is the only necessary ingredient
to reproduce all the relevant features of the Szilard
engine.

Let us remind first what a spontaneous symmetry
breaking is. If ${\cal H}(x)$ is the Hamiltonian
of a system, $x$ being a point in the
phase space $\Gamma$, Statistical Mechanics
prescribes that the probability density for the
equilibrium state of the system at temperature
$T$ is given by the Gibbs distribution:
\begin{equation}
\label{gibbs}
\rho_T(x)={e^{-\beta {\cal H}(x)}\over Z}
\end{equation}
where $\beta=1/kT$ and $Z=\int_\Gamma e^{-\beta{\cal H}}$
is the partition function.
From Eq. (\ref{gibbs})
we see that $\rho_T(x)$ has the same symmetries as
${\cal H}(x)$. Nevertheless, in some cases,
a macroscopic system is not described by the
Gibbs distribution. The phase space
splits into $n$ pieces, $\Gamma_1,\Gamma_2,\dots,
\Gamma_n
\subset\Gamma$ and the macroscopic
system is confined within one of them.
The distribution that describes the system is:
\begin{equation}
\label{gibbsres}
\rho_i(x)={e^{-\beta {\cal H}(x)}\over Z_i}{\cal X}_{\Gamma_i}
(x)
\end{equation}
where ${\cal X}_A(x)$ is the indicator function
of the set $A\subset\Gamma$, i.e., ${\cal X}_A(x)=1$ if $x\in A$
and ${\cal X}_A(x)=0$ if $x\notin A$, and $Z_i$ is
the partition function restricted to $\Gamma_i$.
The distributions $\rho_i(x)$,
called {\em macroscopic phases},
 have less symmetries than
the Hamiltonian. The partition of the phase space, called the {\em
coexistence of macroscopic phases}, occurs for some values of the
temperature and the parameters of the Hamiltonian. Finally, which
of the macroscopic phases is chosen depends on the past of the
system and/or on thermal fluctuations.

The globally coupled
Ising model is one of the simplest systems
exhibiting coexistence of macroscopic phases \cite{huang}.
Its  Hamiltonian is:
\begin{equation}
\label{hamiltonian}
{\cal H} (\{ s_i\};J,B) = -{J\over N} \sum_{i=1}^{N-1}
\sum_{j=i+1}^N s_is_j - B\sum_{i=1}^N s_i
\end{equation}
where the spins take values $s_i=\pm 1$, with $i=1,2,\dots, N$. It
depends on two parameters: the coupling $J$ between spins and the
external field $B$. It is called {\em globally coupled } because
every spin interacts with all the others.

The system exhibits coexistence of two macroscopic
phases when $B=0$ and $J/kT>1$. One of the
phases is restricted to $\Gamma_+$, the
set of configurations $\{ s_i\}$ with
positive global magnetization $M\equiv
\sum_i s_i>0$, and the other is restricted to $\Gamma_-$,
the set of configurations with negative magnetization.
Each phase breaks the symmetry
$\{ s_i\}\to \{ -s_i\}$ that
the Hamiltonian possesses for $B=0$.

When temperature is
lowered,
keeping $B=0$,
 from an initial value above
the critical temperature $T_c\equiv J/k$,
a second
order phase transition occurs at $T=T_c$.
Below $T_c$ the system is in one of the two macroscopic phases.
None of the phases is favored along the process, since
$B=0$. Therefore, the system chooses the
macroscopic phase at random or, more precisely,
the choice is induced by
thermal fluctuations.

The globally coupled Ising
model also exhibits  first order phase transitions
when the field crosses $B=0$ below $T_c$.
The external field breaks the symmetry
$\{ s_i\}\to \{ -s_i\}$ of the Hamiltonian
and, if the coexistence region is reached decreasing
a positive field, the macroscopic phase is
the one with positive magnetization. This can
be called forced or non-spontaneous
symmetry breaking.

To reproduce in the Ising model the two processes, $A$ and $C$,
discussed in Section \ref{sec:2procs} for the Szilard engine, we
need to induce a spontaneous symmetry breaking at constant
temperature (remember that processes $A$ and $C$ in the Szilard
engine are isothermal). This can be achieved if we tune the
coupling $J$ at constant temperature $T$. The spontaneous symmetry
breaking occurs then for a critical coupling $J_c\equiv 1/kT$, and
for $B=0$ and $J>J_c$  the system exhibits coexistence of phases.

\section{Two processes in the
Ising model}
\label{sec:processes}

\begin{figure}[b!]
\centerline{\epsfig{file=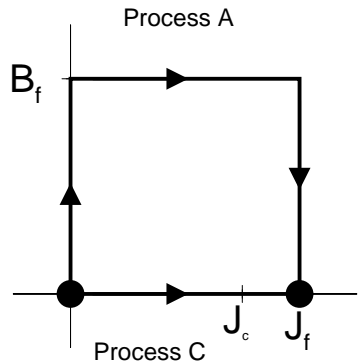,height=5cm}}
\vspace{10pt}
\caption{Processes $A$ and $C$ in the Ising model. The
two black circles are the initial and final states of both
processes.}
\label{figproc}
\end{figure}

Consider the following two processes on the plane
$(J,B)$ (see Figure \ref{figproc}):
\begin{itemize}
\item {\bf Process $A$}: starting at $(0,0)$,
the field is increased
up to $B_f>0$, then the coupling is increased up to
$J_f>J_c$, then the field is decreased down to zero.

\item {\bf Process $C$}: starting at $(0,0)$,
the coupling is increased up to
$J_f>J_c$, keeping $B=0$.

\end{itemize}

The two processes are quasistatic in the following sense: they are
slow enough for the system to relax within {\em each possible
macroscopic phase}, but fast enough for the system to remain in
one of the two phases.

Applying to process $A$
the formalism described in
the Appendix,
one finds the following
energetics, up to order $kT$ \cite{parr}:
\begin{equation}
W_A =
-{\cal F}(T,J_f,0)+
{\cal F}(T,0,0) -kT\ln 2
\label{workA}
\end{equation}
where $ {\cal F}(T,J,B)= -kT\ln Z(\beta,J,B) $ and $
Z(\beta;J,B)=\sum e^{-\beta {\cal H}} $ is the partition function
of the system. $Z(\beta;J,B)$ and ${\cal F}(T,J,B)$ must be
considered here as mere mathematical definitions and we should
refrain from attributing them any physical meaning at this stage
of the discussion. For process $C$ one has:
\begin{equation}
\label{workC}
W_C=- {\cal F}(T,J_f,0)+ {\cal F}(T,0,0).
\end{equation}

Therefore, $W_A-W_C=-kT\ln 2$, i.e., the
external agent has
to do more work to complete process $A$ than to
complete $C$, exactly as in the Szilard
engine.

The whole discussion on the Szilard engine in Sections
\ref{sec:intro} and \ref{sec:2procs} can be applied to the Ising
model. For instance, one can design a cyclic engine as $CA^{-1}$.

Let us first analyze the inverse
processes $A^{-1}$ and $C^{-1}$ in detail.
The inversion of $C$ does not present any difficulty. The
energetics of $C^{-1}$ is
simply $W_{C^{-1}} = -W_{C}$ and
$Q_{C^{-1}} = -Q_{C}$.

\begin{figure}[b!]
\centerline{\epsfig{file=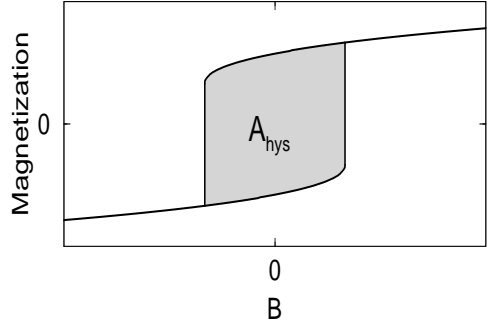,height=4.4cm,width=6.5cm}}
\vspace{10pt}
\caption{Hysteresis cycle in the Ising model.}
\label{fighys}
\end{figure}

On the other hand,
if we try to invert $A$,
{\em the sign of the field must be the same as the sign of the
initial magnetization of the system}.
If we start to increase a positive field on a system with
negative magnetization, the system becomes metastable, it
runs along one of the branches of a hysteresis cycle and
eventually relaxes irreversibly to the stable state for some
value of the field $B$ (see Fig. \ref{fighys}).

The most general case is when we have an ensemble of
systems. If initially
a fraction $\alpha$ of them have negative magnetization,
the energetics of $A^{-1}$ is given by:
\begin{equation}
W_{A^{-1}} =
-W_{A} -\alpha {A_{hys}\over 2}
\label{workA-1}
\end{equation}
where $A_{hys}$ is the area of the hysteresis cycle at
$J=J_f$, as shown in Fig. \ref{fighys}.

The hysteresis phenomenon is not present in
the Szilard engine. However, it has similar consequences
as exerting the pressure in the wrong direction
along the expansion, since in both cases
the system evolves irreversibly doing less work.

Consider now the equivalent to the
Szilard engine, i.e., the cycle $CA^{-1}$
on an
ensemble of Ising models.
Its energetics (per system) is immediately
obtained from Eqs. (\ref{workA}-\ref{workA-1}):
\begin{equation}
W_{CA^{-1}}=W_C+W_{A^{-1}} =kT \ln 2-
\alpha {A_{hys}\over 2}.
\end{equation}
where $\alpha$ is the fraction of systems with magnetization
of the same sign as the field in $A^{-1}$.
There are two consequences of this expression.

First,
if instead of an ensemble we take a single system
and measure its magnetization after $C$
to decide the sign of the field,
then $\alpha=0$ and
 $W_{CA^{-1}}=kT \ln 2>0$, i.e.,
the system is extracting energy from
the thermal bath and converting it into work.
We recover the same result as in the
Szilard engine but now with a genuine macroscopic
system. Thus, we have
a Maxwell demon with the important novelty
that he needs to measure a {\em macroscopic
quantity}.

Second, for an ensemble, $\alpha=1/2$, and
we still can beat the Second Law
unless:
\begin{equation}
\label{hys} A_{hys} \ge 4kT\ln 2.
\end{equation}

This inequality is a byproduct of this
theory and clarifying
its origin  is one of the open problems of the present work.

\section{Entropy and macroscopic uncertainty}
\label{sec:entro}

The above discussion has focused on energy. I will explore in this
Section the consequences of the previous results on the definition
of entropy.

The change of entropy in the thermal bath along
a process
is given by $\Delta S_{bath}=-Q/T$, whereas the
entropy of the external agent is constant because its
interaction with the system is purely mechanical. Then
the change of the total entropy is:
\begin{equation}
\Delta S_{total}=-{Q\over T}+\Delta S_{syst}.
\label{pre2}
\end{equation}
Second Law of Thermodynamics tells us that,
if a process is reversible, $\Delta S_{total} = 0$,
 and, if it is irreversible,
 $\Delta S_{total} > 0$.
In particular,
for a cyclic process, $\Delta S_{syst}=0$ hence $Q\le 0$.
This is the Kelvin-Planck statement of the Second Law: {\em it
is not possible to extract energy from a single thermal
bath in a cyclic process}.

However, Eq. (\ref{pre2}) and the Second Law lead to
contradictions when applied to processes $A$ and $C$. For
instance, $\Delta S_{total}^{CC^{-1}}=\Delta
S_{total}^{AA^{-1}}=0$. Therefore, $AA^{-1}$ and $CC^{-1}$ are
reversible and so are their components, $A$, $A^{-1}$, $C$, and
$C^{-1}$. On the other hand $\Delta S_{total}^{AC^{-1}}= k\ln 2$,
hence $AC^{-1}$ is irreversible. Moreover, if $A$ and $C$ are
reversible, we obtain $\Delta S_{syst}^C= \Delta S_{syst}^A+k\ln
2$.

These contradictions are usually explained
with the following definition for the
thermodynamic entropy of the system:
\begin{equation}
S^{(ens)}_{syst}=-k\langle \ln \rho_{ens}\rangle
\label{subj}
\end{equation}
where $\rho_{ens}$ is the probability distribution describing an
ensemble of systems. After process $C$,
$\rho_{ens}=(\rho_++\rho_-)/2$ where $\rho_+$ and $\rho_-$ are the
probability distribution of the two macroscopic phases (see
Section \ref{sec:sb}). On the other hand, after $A$,
$\rho_{ens}=\rho_+$. Then, $S^{(ens)}_{syst}$ is $k\ln 2$ bigger
after $C$ than after $A$.

This picture is, however, rather unsatisfactory
if we deal with single systems instead with ensembles,
since $\rho_{ens}$ becomes a subjective quantity.
For instance,
the physical state of an Ising model after process $A$ is
the same as after $C$ if the final magnetization is
positive.
The only difference between these two situations
is that we ignore the magnetization after $C$.
Then $S^{(ens)}_{syst}$, as defined in Eq. (\ref{subj}), is a subjective
quantity for single systems.
Mathematically, this can be expressed as:
\begin{equation}
 S^{(ens)}_{syst} =
-k \langle  \ln\rho_{single}\rangle  + H. \label{entro}
\end{equation}
Here, $\rho_{single}$ is the
invariant measure that gives
the temporal average of any magnitude and
it is a fully objective distribution
for a single system \cite{ergo}.
$H$ is the ignorance or uncertainty that we have about
the macrostate of the system.
It is measured using Shannon formula: $H=-k\sum_i
p_i\ln p_i$ where $p_i$ is the probability of having an
instance $i$ (in the Szilard and Ising case, after $C$,
$H=1 \mbox{bit}=k\ln 2$).

Moreover, in this interpretation not only entropy is subjective
but also the concept of reversibility. Consider $C^{-1}$ on a
single system: it is reversible if we do not know the initial
macroscopic magnetization and it is irreversible if we do know it.

I propose a simpler interpretation of the above results. In this
new interpretation, entropy is an objective magnitude for single
systems, but we are forced to admit that it decreases along some
processes, in contradiction with {\em some} formulations of the
Second Law.  However, the main limitation imposed by the Second
Law, namely, the Kelvin-Planck statement, remains valid, since
these processes cannot be used to construct cycles. Ishioka and
Fuchikami, in these proceedings \cite{ishioka}, have reached
similar conclussions. The assumptions for this interpretation are
the following:
\begin{enumerate}
\item The thermodynamic entropy of a system is given by:
\begin{equation}
S_{syst} \equiv
 -k\langle\ln \rho_{single}\rangle.
\label{sobj}
\end{equation}
\item If an external agent induces,
in a quasistatic and isothermal way, a spontaneous symmetry
breaking with $n$ phases, the total entropy (the sum of the
entropies of the system, thermal bath, and external agent)
decreases by $k\ln n$. I will call these processes {\bf
anti-irreversible} (in \cite{ishioka} the term {\em partitioning
processes} is used instead) and they correspond to the {\em
creation of macroscopic uncertainty}.
\item Along the inverse of an anti-irreversible
process, the total entropy increases by $k\ln n$.
I will call these processes {\bf quasi-irreversible} or simply
irreversible.
\end{enumerate}
Process $C$ is anti-irreversible and $C^{-1}$ is
quasi-irreversible. The reason of the names is the following:
$C^{-1}$ cannot be truly reversed because, after $C^{-1}C$ the
initial magnetization could be opposite to the final one.
Processes $A$ and $A^{-1}$ are reversible in the standard sense,
i.e., total entropy does not change. The reader can check that
every combination of processes $A$, $C$ and their inversions are
explained with the above three rules. Moreover, entropy and
reversibility become fully objective concepts.

The measurement process can be also
explained with this new Thermodynamics.
Consider, as a model of a system
and a measurement device, the
Hamiltonian:
\[
{\cal H} (\{ s_i^{(1)}\},\{s_i^{(2)}\};J_1,J_2,J_{12}) =
-{J_1\over N} \sum_{j>i}^{N}s^{(1)}_is^{(1)}_j
-{J_2\over N} \sum_{j>i}^{N} s^{(2)}_is^{(2)}_j
-{J_{12}\over N} \sum_{i,j=1}^{N}s^{(1)}_is^{(2)}_j
\]
which corresponds to two  coupled Ising models, 1 (system) and 2
(measurement device or `pointer'). The following table shows the
behavior of the total entropy, as defined by (\ref{pre2}) and
(\ref{sobj}), and the macroscopic uncertainty $H$, along two
isothermal and quasistatic processes: \vspace{.3cm}

\begin{center}
\begin{tabular}{rcccc|cccc}
&Step & $S_{total}-S_{total}^0$
&  $H$
 &
\hspace{.1cm}  &
\hspace{.1cm}  &
Step & $S_{total}-S_{total}^0$
& $H$
 \\
\cline{1-4} \cline{7-9}
 &&&&&&&& \\
1)& $J_1:0\to J_f$ & $-k\ln 2$ & 1 bit  &&
& $J_1:0\to J_f$ & $-k\ln 2$ & 1 bit
 \\
2)& $J_{12}:0\to J_f$ & $-k\ln 2$ & 1 bit    & &
&  $J_{12}:0\to J_f$ & $-k\ln 2$  & 1 bit    \\
3)& $J_2:0\to J_f$ & $-k\ln 2$  & 1 bit    &&
& $J_2:0\to J_f$ & $-k\ln 2$  & 1 bit
  \\
4)& $J_{1}:J_f\to 0$ & $-k\ln 2$  & 1 bit    &&
& $J_{12}:J_f\to 0$ & $-k\ln 2$  & 1 bit
  \\
5)& $J_{12}:J_f\to 0$ & $-k\ln 2$  & 1 bit    &&
& $J_{1}:J_f\to 0$ & $0$  & 1 bit
  \\
6)& $J_{2}:J_f\to 0$ & $0$  & $0$ &&
& $J_{2}:J_f\to 0$ & $k\ln 2$  & $0$
  \\
\end{tabular}
\end{center}
\vspace{.3cm}

Both processes
involve a spontaneous symmetry breaking (step 1),
copying the outcome (steps 2-3)
and erase of the copy and the original (steps 4-6).

The first process (left column) can be interpreted as a reversible
measurement. Measurement can be defined in a rather general way as
any procedure which allows one to drive a system from the region
of coexistence of phases to a region of non-coexistence along a
reversible process, i.e., avoiding the critical point as well as
the possibility of hysteresis. This is done in step 4 of the first
column, where $J_1$ is decreased down to zero along a reversible
process. As a result, the total entropy is lowered  by $k\ln 2$ in
the first five steps. Notice also that, to drive the whole system
1+2 to its initial state, we have to {\em reset} the {\em
measurement device} 2, by crossing again a critical point, i.e.,
along a quasi-irreversible process (step 6). We thus recover
Bennet's interpretation of the Szilard engine \cite{bennet}.

I have included the other process (right column in
the table) to show how subtle
the measurement and the erasure processes can be. If
subsystem 1 is uncoupled before driven to its
initial state, then it crosses a critical
point and the entropy increases. Step 5 in
the right column is quasi-irreversible, because initially
the magnetizations of
1 and 2 have the same sign, and, if step 5 were reversed
the final magnetizations would be uncorrelated.
A similar effect of the correlation between the
particle and the measurement device
in the Szilard engine was pointed out
by Fahn in Ref. \cite{fahn}.

\section{Conclusions and open problems}
\label{sec:open}

Two objections can be raised against the Thermodynamics proposed
in the last Section. The first is that energy is an extensive
property, i.e., of order $NkT$, and terms of order $kT\ln 2$ are
negligible and even  much smaller than the energy fluctuations.
This objection applies to any Maxwell demon but it is not
sufficient to exorcize it. The reason is that the demon can repeat
the cycle as many times as he wants, converting a macroscopic
amount of heat into work.

The second objection is that the increase of
entropy can be derived from non-equilibrium theories,
such as the Fokker-Planck formalism.
If
${\bf q}$ are  the (overdamped)
degrees of freedom of a system,
the probability distribution obeys the Fokker-Planck equation
(FPE):
\begin{equation}
\partial_t \rho({\bf q},t) = -\nabla \cdot{\bf J}({\bf q},t)
\label{fp}
\end{equation}
where the current is ${\bf J}({\bf q},t)=[
-\nabla \mu ({\bf q},t)]\rho({\bf q},t)$ and
 the chemical potential is defined as
$\mu({\bf q},t) \equiv V( {\bf q},t)+kT
\ln\rho({\bf q},t)$.
From these equations one can derive the following identity
\cite{magszil}:
\begin{eqnarray}
-k\partial_t \int d{\bf q}
\ \rho({\bf q},t)
\ln\rho({\bf q},t)
 &=&
{1\over T}
 \int d{\bf q}
V ({\bf q},t)\partial_t
\rho({\bf q},t)
+
{1\over T}
 \int d{\bf q}
{|{\bf J}({\bf q},t)|^2\over
\rho({\bf q},t) }
\nonumber
\\
&=&
 {\dot Q\over T} +
{1\over T}
 \int d{\bf q}
{|{\bf J}({\bf q},t)|^2\over
\rho({\bf q},t) }
\label{fp2}
 \end{eqnarray}

If the left-hand side of this equation is interpreted as $\dot
S_{syst}$, the change of the entropy of the system per unit of
time, then the total change of entropy, $\dot S_{total} = -\dot
Q/T+ \dot S_{syst}$, is always positive. A similar result can be
obtained for underdamped degrees of freedom \cite{shizume}. How
then have we obtained $\dot S_{total}<0$ for some processes
involving phase transitions? The answer is that the distribution
that appears in the FPE (\ref{fp}) is $\rho_{ens}$ and not
$\rho_{single}$. Then, the FPE is not appropriate to describe
single macroscopic systems.

One of the open problems of the present work is to characterize
$\rho_{single}$  and derive an evolution equation similar to the
FPE. Other open problems are: (a) analyze the role of hysteresis
and the origin of inequality (\ref{hys}); (b) extend the above
discussion to the breaking of a continuous symmetry, where an
infinite number of macroscopic phases coexist; (c) include the
external agent in the Hamiltonian as a set of macroscopic degrees
of freedom; and (d) explore the implications of the decrease of
entropy along anti-irreversible processes, specially in cosmology.

\section*{Appendix}

Consider a system whose
Hamiltonian
${\cal H}(x;{\bf R})$
depends on
a set of external parameters
collected in a vector ${\bf R}$.
We are interested in the energetics of a
process along which
the system is in contact with a thermal bath
at temperature $T$ and
the parameters are changed by an external
agent as ${\bf R}(t)$ with $t\in [0,{\cal T}]$.

The expressions for work and heat per unit of time along this
process are \cite{shizume,denbigh}:
\begin{equation}
\dot Q =
\int_\Gamma dx {\cal H}( x;{\bf R}(t)){\partial \rho(x;t)\over
\partial t}\ ;\qquad
\dot W =-
\int_\Gamma dx
\rho(x;t)
{\partial
 {\cal H}(x;{\bf R}(t))
\over \partial t}
\label{hw1}
\end{equation}
where $\Gamma$ is the phase space of the system and
$\rho(x;t)$ the probability density for
the state $x$.
Heat and work, as given by
Eq. (\ref{hw1}),
obey the First Law of
Thermodynamics: $\dot {\cal U} = \dot Q-\dot W$.

If the process is quasistatic, ${\cal T}\to\infty$,
the probability
density
 at time $t$ depends only on the value of
the external parameters at $t$, i.e.,
$\rho(x;t)=\rho(x;{\bf R}(t))$. In this case, the heat
and the work in the whole process are given by:
\begin{equation}
Q=\int_A \delta Q({\bf R})\ ;
\qquad
\qquad
W=\int_A \delta W({\bf R})
\label{hw1bis}
\end{equation}
where $A$ is the path that ${\bf R}(t)$ describes
along the process and the infinitesimal work and heat
are given by:
\begin{eqnarray}
\delta Q ({\bf R})&=&
\int_\Gamma dx\ {\cal H}( x;{\bf R}){\partial \rho(x;{\bf R})\over
\partial {\bf R}}\cdot d{\bf R}
\nonumber \\
\delta W ({\bf R})&=&-
\int_\Gamma dx\
\rho(x;{\bf R})
{\partial
 {\cal H}(x;{\bf R})
\over \partial {\bf R}}\cdot d{\bf R}.
\label{hw2}
\end{eqnarray}

The most familiar implementation of the above expressions
is obtained when the  state of the system is the Gibbs
distribution,  $\rho_T(x;{\bf R})\equiv
e^{-\beta {\cal H}(x;{\bf R})}/ Z(\beta,{\bf R})$.
For this particular case,
Eqs. (\ref{hw2}) reduce to
\begin{equation}
\delta Q({\bf R}) =
T{\partial S(T,{\bf R})\over \partial {\bf R}}\cdot
d{\bf R}\ ;
\qquad
\delta W ({\bf R}) = -
{\partial {\cal F}(T,{\bf R})\over \partial {\bf R}}\cdot
d{\bf R}
\label{standard}
\end{equation}
where
$ S(T,{\bf R})  =-k\int_\Gamma dx\
\rho_{T}(x;{\bf R})
\ln [\rho_{T}(x;{\bf R})]$
and ${\cal F}(T,{\bf R})= -kT\ln Z(\beta,{\bf R})$
are, respectively, the free energy and the
entropy of the system.

Although processes $A$ and $C$
considered in the text are
isothermal  and quasistatic,
the state $\rho(x;{\bf R})$ is not equal to
$\rho_T(x;{\bf R})$
in the region of coexistence of macroscopic phases.
Consequently
their energetics, up to terms
of order $kT$, differ from the one prescribed by
standard equilibrium Thermodynamics.

To get Eqs.\ (\ref{wac}), I have used
Eqs.\ (\ref{hw1bis}) and (\ref{hw2}) with the following
prescription for $\rho (\{ s_i\};J,B)$
along process $C$:
$\rho(\{ s_i\}; J,0) =
\rho_T(\{ s_i\}; J,B)$ if $J<J_c$ and
$\rho(\{ s_i\}; J,0) =
\rho_+(\{ s_i\}; J,B)$ if $J\ge J_c$,
where $\rho_+$ is $\rho_T$ restricted to $\Gamma_+$,
the set of configurations with positive magnetization.
The precise location
of the replacement of $\rho_T$ by $\rho_+$ does
not affect the results.
In fact, the energetics is the same as if calculated
using $\rho_T$, for symmetry reasons \cite{parr}.

Along process $A$, the state is given by:
$\rho_T(\{ s_i\}; J,B)$ if $J=0$ or $B=B_f$ (first
two steps of $A$)
and by
$\rho_+(\{ s_i\}; J,B)$ if $J= J_f$ (last step).
Again   the
energetics, up to order $kT$, does
not depend on where precisely the system changes from
$\rho_T$ to $\rho_+$. The above prescription has been chosen
for simplicity. The replacement of $\rho_T$ by
$\rho_+$ is only significant at the end of
the last step, i.e.,
when the system is close to
the region of coexistence.
$W_A$ can be calculated by using
(\ref{standard}) along the first two steps and
using the partition function restricted to $\Gamma_+$ along
the third step. It can be rigorously proven that
the energetics is the one given by (\ref{wac}) plus
terms of order $kTe^{-\gamma N}$, where $\gamma$ is positive
and  of order $o(1)$ if $B_f$ and $J_f$ are
large enough. Details of the calculations will be given
 elsewhere \cite{parr}.

\end{document}